\documentclass[envcountsame]{llncs}
\usepackage{amsmath,amsfonts,amssymb}
\usepackage{multicol}
\usepackage{hyperref}
\usepackage{cleveref}
\usepackage{url}
\usepackage{color}
\usepackage{mathpartir}
\usepackage{centernot}
\usepackage[mathcal]{euscript}

\newcommand{\CSPe}{\ensuremath{\text{CSP}_E}}

\usepackage{ifthen}
\newcommand{\ifempty}[3]{\ifthenelse{\equal{#1}{}}{#2}{#3}}

\newcommand{\tick}{\checkmark}
\usepackage{stmaryrd}

\newcommand{\parcomp}[1][E]{%
  \mathrel{||_{#1}}%
}
\newcommand{\xchoice}{\mathrel{\Box}}
\newcommand{\FAIL}{\textit{FAIL}}

\newcommand{\STOP}{\textit{STOP}}
\newcommand{\sem}[1]{\llbracket #1 \rrbracket}

\newcommand{\sstep}[1]{\mathrel{\overset{#1}{\mapsto}}}
\newcommand{\ssteps}[1]{\mathrel{\overset{#1}{\Mapsto}}}

\newcommand{\Doomed}{\textit{Doomed}}
\newcommand{\Terms}{\textit{Terms}}
\newcommand{\viable}[1]{\hat{#1}}

\newcommand{\TraceSets}{\textit{TraceSets}}

\newif{\ifjournal}
\journalfalse

\begin{document}
\title{Operational Semantics of Process Monitors}

\author{Jun Inoue and Yoriyuki Yamagata}

\institute{National Institute of Advanced Industrial Science and 
  Technology (AIST), 1-8-31 Midorigaoka, Ikeda, Osaka 563-8577 Japan
  \email{\{jun.inoue, yoriyuki.yamagata\}@aist.go.jp}
}
\maketitle
\begin{abstract}
  \CSPe{} is a specification language for runtime monitors that can 
  directly express concurrency in a bottom-up manner that composes 
  the system from simpler, interacting components.
  It includes constructs to explicitly flag failures to the monitor, 
  which unlike deadlocks and livelocks in conventional process 
  algebras, propagate globally and aborts the whole system's 
  execution.
  Although \CSPe{} has a trace semantics along with an implementation 
  demonstrating acceptable performance, it lacks an operational 
  semantics.
  An operational semantics is not only more accessible than trace 
  semantics but also indispensable for ensuring the correctness of 
  the implementation.
  Furthermore, a process algebra like \CSPe{} admits multiple 
  denotational semantics appropriate for different purposes, and an 
  operational semantics is the basis for justifying such semantics' 
  integrity and relevance.
  In this paper, we develop an SOS-style operational semantics for 
  \CSPe{}, which properly accounts for explicit failures and will 
  serve as a basis for further study of its properties, its 
  optimization, and its use in runtime verification.
\end{abstract}


\section{Introduction}
\label{sec:intro}

Specification-based runtime monitoring 
\cite{Havelund2015spec-param-mon} checks a program's execution trace 
against a formal specification.
Often more rigorous than testing due to the presence of a formal 
specification, this technique is also computationally much cheaper 
than formal verification methods like model checking, as it only 
needs to look at concrete program runs with instrumentation.
\CSPe{} \cite{Yamagata16runtime-monitoring} is a language based on 
Hoare's Communicating Sequential Processes \cite{Hoare85csp} for 
developing the formal specification.
Unlike many other languages in this niche, \CSPe{} can directly 
express concurrency.
Moreover, it builds up the specification in a bottom-up manner by 
composing smaller, interacting components, helping to model complex 
behavior.

\CSPe{}'s main appeal as a specification language, compared to plain 
CSP, is a $\FAIL$ construct that signals a global failure, aborting 
all processes in the model at once.
This construct can be used like \texttt{assert(false)} in C or Java, 
allowing to code invariants that mark some states as (should-be) 
unreachable.
By contrast, deadlocks and livelocks, the conventional notions of 
failure in CSP, affect only the deadlocked or livelocked process(es).
These failures are thus very difficult to propagate into a failure 
for the entire system, as desired for assertion failures.
However, the semantics of how $\FAIL$ propagates throughout the model 
requires special treatment.
Because the propagation preempts all other activities, normal 
execution rules must apply only when $\FAIL$ is not currently 
propagating.
This is a negative constraint, which is generally problematic 
\cite{Groote93tss-with-negative}.

While earlier work \cite{Yamagata16runtime-monitoring} demonstrated a 
trace semantics and a reasonably efficient implementation for 
\CSPe{}, an operational semantics has been lacking.
Developing an operational semantics is highly desirable for several 
reasons.
Firstly, though a trace semantics more naturally defines the set of 
behaviors (i.e.\ traces) that comply with a \CSPe{} specification, an 
operational semantics more directly defines the implementation.
Secondly, process algebras admit multiple denotational semantics 
capturing different aspects of operationally defined behavior 
\cite{vanGlabbeek01spectrum}.
Investigating the full spectrum of such semantics requires an 
operational semantics.
Finally, an operational semantics provides a more accessible 
presentation of the semantics than denotational semantics.

\subsection{Contributions}
\label{sec:contrib}

In this paper, after reviewing the syntax and trace semantics of 
\CSPe{} (\Cref{sec:syntax}), we present the following contributions.
\begin{itemize}
\item We define an operational semantics in SOS format 
  \cite{Plotkin04sos}, which properly captures the propagation of 
  $\FAIL$ while avoiding the complexities of rules with negative 
  premises (\Cref{sec:op-sem}).
\item We prove that the operational semantics induces the previously 
  published trace semantics (\Cref{sec:correspondence}).
\end{itemize}

\section{Syntax and Trace Semantics of \CSPe{}}
\label{sec:syntax}

\begin{figure}[t]
  \centering
  \begin{tabular}{@{}l@{\hspace{1em}}l@{}}
    \emph{Event} & $e \in \Sigma$ \\
    \emph{Event Variable} & $x \in X$ \\
    \emph{Term} & $P,Q \in \Terms ::= \STOP \mid \FAIL \mid \phantom{}?x:E \rightarrow P \mid P \xchoice Q \mid P \parcomp Q$ \\ 
    \emph{Event Set} & $E ::= f(y_1, \ldots, y_n)$ where $f : \Sigma^n \rightarrow 2^\Sigma$ is computable \\ 
    \emph{Event Set Param} & $y ::= x \mid e$
  \end{tabular}
  \caption{Syntax of \CSPe{}.}
  \label{fig:syntax}
\end{figure}

This section reviews the syntax and trace semantics of \CSPe{}.
\Cref{fig:syntax} presents the syntax.
A \CSPe{} term represents a process, which is an entity that 
successively emits \emph{events} drawn from an alphabet $\Sigma$.
Terms are built from %
\iffalse%
the terminating term, the deadlocked term, the failing term, prefix 
(which emits one event chosen from an event set), sequencing, 
external choice, and parallel composition with synchronization on an 
event set.
Prefix $?x:E \rightarrow P$ and parallel composition $P \parcomp Q$ 
take a set of events $E$, which is the set of events that can be 
emitted in the case of prefix, the following constructs, with the 
following intuitive meanings.
\else%
the following constructs, with the indicated meanings.
For a thorougher explanation, see 
\cite{Yamagata16runtime-monitoring}.
\begin{itemize}
\item The stuck term $\STOP$ does not emit anything.
\item The failing term $\FAIL$ aborts all processes.
\item Prefix $?x:E \rightarrow P$ chooses and emits an event 
  $e \in E$, then executes $[e/x]P$.
\item Choice $P \xchoice Q$ executes $P$ or $Q$, whichever manages to 
  emit something first.
\item Parallel composition $P \parcomp Q$ executes $P$ and $Q$ in 
  parallel.
  Their events are interleaved arbitrarily, except events in $E$ are 
  synchronized.
\end{itemize}
\fi%
An event set $E$ can be specified by any computable function 
parametrized by the $x$'s bound by surrounding prefix operators.

In this short paper, we omit recursion and the terminating action 
$\tick$ in the interest of conciseness.
This paper's focus is on analyzing $\FAIL$, and $\tick$ complicates 
the presentation substantially without adding anything of conceptual 
significance.
Recursion seems to be similar, though it is still under 
investigation.

\begin{figure}[t]
  \centering
  \begin{tabular}{@{}l@{\hspace{1em}}l@{}}
    \emph{Trace} & $s,t \in \Sigma^*$ \\ 
    \emph{Trace Set} & $T \in \TraceSets ::= \text{prefix-closed subsets of }\Sigma^*$ \\
    \emph{Trace Set Operations} \\
    \multicolumn{2}{@{}l@{}}{%
    \begin{minipage}{1.0\linewidth}
      \vspace{-0.7\baselineskip}
      \begin{align*}
        eT & := \{\varepsilon\}\cup \{et \mid t \in T\} \\
        T(e) & := \{t \mid et \in T\} \\
        \varnothing \parcomp T & := T \parcomp \varnothing := \varnothing
                                 \stepcounter{equation}\tag{\theequation}\label{eq:parcomp:empty} \\
        T_1 \parcomp T_2 & := \bigcup_{e \in E}e(T_1(e) \parcomp T_2(e)) \cup \bigcup_{e \in \Sigma-E}\left(e(T_1(e) \parcomp T_2) \cup e(T_1 \parcomp T_2(e))\right)
             \label{eq:parcomp:nonempty}\stepcounter{equation}\tag{\theequation}
      \end{align*}
    \end{minipage}
          } \\
    \emph{Trace Semantics} \\
    \multicolumn{2}{@{}l@{}}{
    \begin{minipage}{1.0\linewidth}
      \vspace{-0.7\baselineskip}
      \begin{align*}
        \sem{\STOP} & := \{\varepsilon\} & \sem{P \xchoice Q} & := \sem{P} \cup \sem{Q} \\
        \sem{\FAIL} & := \varnothing & \sem{P \parcomp Q} & := \sem{P} \parcomp \sem{Q} \\
        \sem{?x:E \rightarrow P} & := \{\varepsilon\} \cup \bigcup_{e \in E}e\sem{[e/x]P}
      \end{align*}
    \end{minipage}
             } \\
  \end{tabular}
  \caption{Trace semantics of \CSPe{}.
    \Cref{eq:parcomp:empty} takes precedence over 
    \cref{eq:parcomp:nonempty}, so the latter applies only if the 
    former does not.}
  \label{fig:trace-sem}
\end{figure}

\Cref{fig:trace-sem} presents the trace semantics.
A \emph{trace} is a (possibly empty) sequence of events, and 
$\Sigma^*$ is the set of all traces.
The concatenation of traces $s$ and $t$ is written $st$.
A \emph{trace set} is any prefix-closed set of traces, which can be 
empty, unlike in conventional process algebras.
The trace semantics of \CSPe{} assigns to each term $P$ a trace set 
$\sem{P}$, which is intuitively the set of traces $P$ can emit.

The semantic map uses some operations on trace sets.
If $T$ is a trace set, $eT$ prepends $e$ to all members of $T$ and 
adjoins $\varepsilon$, while $T(e)$ discards all traces in $T$ that 
do not start with $e$ and drops the leading $e$ from all remaining 
traces.
The $\parcomp$ operator is defined by 
\cref{eq:parcomp:empty,eq:parcomp:nonempty}.
Though significantly simplified, these equations are equivalent to 
the ones found in \cite{Yamagata16runtime-monitoring} modulo the 
absence of $\tick$.
In \cite{Yamagata16runtime-monitoring}, this operator was defined 
``coinductively'', which was correct but misleading.
Formally, by the Knaster-Tarski Theorem, the defining equations 
\eqref{eq:parcomp:empty} and \eqref{eq:parcomp:nonempty} have a 
greatest solution in the complete lattice of total binary functions 
on $\TraceSets$ ordered by point-wise inclusion, which was taken to 
be $\parcomp$.
However, if $\parcomp'$ and $\parcomp''$ are any two solutions of 
these equations, then for any $T_1$ and $T_2$, every trace in 
$T_1 \parcomp' T_2$ is also in $T_1 \parcomp'' T_2$, by 
straightforward induction on the trace's length.
Thus, the solution is unique, and $\parcomp$ is this unique solution.

\begin{lemma}\label{thm:parcomp-cont}
  $\parcomp$ is continuous, i.e.\ 
  $\left(\bigcup S_1\right) \parcomp \left(\bigcup 
    S_2\right) = \bigcup_{T_1 \in S_1, T_2 \in S_2} T_1 \parcomp 
  T_2$.
  \begin{proof}
    The defining equations \eqref{eq:parcomp:empty} and 
    \eqref{eq:parcomp:nonempty} preserve continuity, so in fact the 
    Knaster-Tarski construction can be carried out in the space of 
    continuous binary operators, which is also a complete lattice 
    under point-wise inclusion.
  \end{proof}
\end{lemma}

\section{Operational Semantics}
\label{sec:op-sem}

\begin{figure}[t]
  \centering
  \begin{tabular}{@{}l@{\hspace{1em}}l@{}}
    \emph{Action} & $a ::= e \mid \tau$ \\
    \emph{Doomed Term} & $D \in \Doomed ::= \FAIL \mid D \xchoice D \mid D \parcomp P \mid P \parcomp D$ \\
    \emph{Viable Term} & $\viable{P},\viable{Q} \in \Terms - \Doomed$ \\
    \emph{Operational Semantics} \\
    \multicolumn{2}{@{}l@{}}{
    \begin{minipage}{1.0\linewidth}
      \begin{mathpar}
        \inferrule{e \in E}{(?x:E \rightarrow P) \sstep{e} [e/x]P}
        \and
        \inferrule{P \sstep{\tau} P'}{P\xchoice Q \sstep{\tau} P'\xchoice Q}
        \and
        \inferrule{Q \sstep{\tau} Q'}{P\xchoice Q \sstep{\tau} P\xchoice Q'}
        \and
        \inferrule{P \sstep{e} P'}{P\xchoice Q \sstep{e} P'}
        \and
        \inferrule{Q \sstep{e} Q'}{P\xchoice Q \sstep{e} Q'}
        \and
        \inferrule{P \sstep{a} P' \and a \not\in E}{P\parcomp \viable{Q} \sstep{a} P' \parcomp \viable{Q}}
        \and
        \mprset{sep=1.5em}
        \inferrule{Q \sstep{a} Q' \\ a \not\in E}{\viable{P}\parcomp Q \sstep{a} \viable{P} \parcomp Q'}
        \and
        \inferrule{\viable{P} \sstep{e} P' \\ \viable{Q} \sstep{e} Q' \\ e \in E}{\viable{P}\parcomp \viable{Q} \sstep{e} P' \parcomp Q'}
        \and
        \inferrule{D_1 \sstep{\tau} P_1}{D_1\parcomp D_2 \sstep{\tau} P_1 \parcomp D_2}
        \and
        \inferrule{D_2 \sstep{\tau} P_2}{D_1\parcomp D_2 \sstep{\tau} D_1 \parcomp P_2}
        \and
        \inferrule{ }{\FAIL \xchoice \FAIL \sstep{\tau} \FAIL}
        \and
        \inferrule{ }{\FAIL \parcomp P \sstep{\tau} \FAIL}
        \and
        \inferrule{ }{P \parcomp \FAIL \sstep{\tau} \FAIL}
        \and
        \inferrule{ }{P \ssteps{\varepsilon} P}
        \and
        \inferrule{P \sstep{a} P' \ssteps{s} P''}{P\ssteps{as} P''}
        \and
        \inferrule{P \sstep{\tau} P' \ssteps{s} P''}{P \ssteps{s} P''}
      \end{mathpar}
    \end{minipage}
             } \\
  \end{tabular}
  \caption{Operational semantics of \CSPe{}.}
  \label{fig:op-sem}
\end{figure}

This section presents the operational semantics.
The semantics is given in \Cref{fig:op-sem}, which defines 
\emph{internal transitions} $P \sstep{a} Q$ between terms.
Some transitions do not emit events but instead emit the silent 
action $\tau$.
A \emph{visible transition} $P \ssteps{s} Q$ happens when $P$ 
internally transitions to $Q$ in zero or more steps, and the 
non-$\tau$ actions it emits along the way forms $s$.

The main challenge in this semantics is capturing the propagation of 
$\FAIL$.
For example, in $P \parcomp[\varnothing] \FAIL$, the $P$ must not be 
allowed to keep emitting events, for then $P$ could do so 
indefinitely, withholding the propagation of $\FAIL$.
Instead, $\FAIL$ should kill all processes including $P$, 
transitioning the whole term to $\FAIL$.
To achieve this effect, the usual rule that allows the left operand 
to transition must apply only when the right operand is not failing.
This constraint is tricky to capture because it is a negative 
constraint.

In our semantics, the constraint is captured by the viability 
annotation $\viable{P}$.
This annotation restricts the range of the metavariable $\viable{P}$ 
to exclude \emph{doomed} terms, i.e.\ terms for which transitioning 
to $\FAIL$ has become inevitable and are now propagating $\FAIL$ 
within themselves.
These annotations are placed so that when a term is doomed, rules 
that propagate $\FAIL$ become the only applicable ones, thus forcing 
the propagation to take place.

\begin{proposition}\label{thm:doomed-normalizing}
  A doomed process always transitions to $\FAIL$ while emitting 
  nothing but $\tau$.
  \begin{proof}
    $D \sstep{a} P$ implies 
    $a = \tau \wedge P \in \Doomed\wedge |D| > |P|$, where $|P|$ 
    denotes term size, by induction on $D$.
    Thus, a doomed term can only $\tau$-transition, and only finitely 
    many times, while staying doomed.
    Another induction shows 
    $\forall D \neq \FAIL.\ \exists D'.\ D \sstep{\tau} D'$, so a 
    doomed term keeps transitioning until it reaches $\FAIL$.
  \end{proof}
\end{proposition}

\section{Correspondence Between the Semantics}
\label{sec:correspondence}

This section establishes a correspondence between the two semantics: 
a process' denotation is precisely the set of traces it can emit, up 
to but not including any transitions that doom the process.
This means that the monitor comparing a system to $P$ can declare a 
failure as soon as the system's trace strays out of $\sem{P}$.

\begin{theorem}\label{thm:correspondence}
  $\sem{P} = \{s \mid \exists M.\ P \ssteps{s} M \not\in \Doomed\}$.
\end{theorem}

A special case of this theorem is particularly illuminating: the 
doomed set is precisely the set of terms with empty trace sets, 
corresponding to the fact that doomed terms silently transition to 
$\FAIL$.

\begin{proposition}\label{thm:doomed=empty}
  $P \in \Doomed \Longleftrightarrow \sem{P} = \varnothing$.
  \begin{proof}
    Induction on $P$.
  \end{proof}
\end{proposition}
Furthermore, trace sets faithfully follow non-silent transitions, in 
that the traces which follow an event $e$ in $\sem{P}$ are precisely 
the traces of terms $Q$ that follow $P$ after a sequence of 
transitions that emit $e$.
\begin{lemma}\label{thm:sem-visible-trans}
  $\sem{P}(e) = \bigcup_{P \ssteps{e} Q}\sem{Q}$.
  \begin{proof}
    Induction on the size of $P$, where event sets do not count 
    toward size, e.g.\ $|?x:E \rightarrow P'| := |P'| + 1$.
    This way, $|[e/x]P'| = |P'|$, so when 
    $P = (?x:E \rightarrow P')$, the inductive hypothesis applies to 
    $[e/x]P'$, despite it not being a subterm.
    Several lemmas are needed along the way, two of which are of 
    particular note.
    Take for example $P = P_1 \parcomp P_2$ with 
    $\sem{P_1},\sem{P_2} \neq \varnothing$ and $e \in E$.
    Inductive hypotheses give 
    $\sem{P}(e) = (\bigcup_{P_1 \ssteps{e} Q_1}\sem{Q_1})\parcomp 
    (\bigcup_{P_2 \ssteps{e} Q_2}\sem{Q_2})$.
    Then, continuity (\Cref{thm:parcomp-cont}) lets us commute the 
    $\bigcup$ and $\parcomp$, equating this to 
    $\bigcup_{P_1 \ssteps{e} Q_1, P_2 
      \ssteps{e}Q_2}(\sem{Q_1} \parcomp \sem{Q_2})$.
    Then, a lemma characterizing those $Q$ with $P \ssteps{e} Q$ 
    equates this to $\bigcup_{P \ssteps{e} Q}\sem{Q}$.
  \end{proof}
\end{lemma}

\Cref{thm:correspondence} is a straightforward consequence of these 
facts.
\begin{proof}[of \Cref{thm:correspondence}]
  \it We show 
  $s \in \sem{P} \Longleftrightarrow \exists Q.\ P \ssteps{s} Q 
  \not\in \Doomed$ by induction on $s$.
  For the base case, 
  $\varepsilon \in \sem{P} \Longleftrightarrow P \not\in \Doomed$ by 
  \Cref{thm:doomed=empty}.
  If $P \not\in \Doomed$, then 
  $P \ssteps{\varepsilon} P \not\in \Doomed$, and if $P \in \Doomed$, 
  then $P$ can only transition inside $\Doomed$ as noted in the proof 
  of \Cref{thm:doomed-normalizing}.
  For the inductive step, $s$ breaks down as $s = es'$, and 
  $s \in \sem{P} \Longleftrightarrow s' \in \sem{P}(e)$.
  By \Cref{thm:sem-visible-trans}, this is equivalent to having 
  $s' \in \sem{P'}$ and $P \ssteps{e} P'$ for some $P'$, which by 
  inductive hypothesis is equivalent to 
  $\exists P',Q.\ P \ssteps{e} P' \ssteps{s'} Q \not\in \Doomed$.
\end{proof}

\section{Related Works}
\label{sec:related-work}

The main issue with \CSPe{} semantics is the propagation of $\FAIL$, 
which entails the negative constraint that normal computation rules 
apply only if $\FAIL$-propagation rules do not.
Negative premises of the form $P \not\sstep{a}$ come quite naturally 
as a means for codifying such constraints, but negative premises are 
generally quite problematic.
A transition relation satisfying negative rules may be not-existent, 
or non-unique, with no obvious guiding principle (such as minimality) 
in choosing the ``right'' one.
Some formats do guarantee well-definedness, such as GSOS with the 
witnessing constraint \cite{Bloom95bisim-cant-be-traced} and 
\textit{ntyft/ntyxt} \cite{Groote93tss-with-negative}.
But even then, negative rules tend to betray desirable properties 
such as compositionality of some forms of bisimulation 
\cite{Bloom95sos-for-weak}.

Our approach exploits the fact that we only have a very specific 
negative constraint -- the absence of doomed subprocesses -- and 
encodes it with a restriction on the range of metavariables in 
transition rules.
With trick, we manage to avoid negative premises altogether, 
essentially turning the system into a positive one.
This approach is very commonly employed, e.g.\ in reduction rules for 
the call-by-value $\lambda$ calculus \cite{Mitchell96foundations}, 
where the argument in a function application should be evaluated only 
if the function expression cannot be evaluated any further.

We identify $\FAIL$-induced failures by transitions into $\FAIL$, but 
an alternative approach would be to have $\FAIL$ emit a special event 
$\digamma$, just as termination is signalled by $\tick$.  
Though we have not pursued this idea in detail, the central concern 
there will be to give $\digamma$ higher priority than all other 
events.
Prioritized transition also involves a negative constraint but is 
known to be quite well-behaved, being translatable to plain CSP 
\cite{Roscoe15csp-pri}.
At the moment, it is not clear if $\FAIL$ propagation can be 
translated to the prioritized-transition primitive in 
\cite{Roscoe15csp-pri}.

\section{Conclusion}

We gave an operational semantics for \CSPe{} that adequately captures 
the behavior of $\FAIL$, the global failure operator, with positive 
operational rules.
This semantics induces the previously defined trace semantics.
As noted in the introduction, this development enables studies of 
other types of denotational semantics, while informing the 
implementation.
An interesting direction of future work is to see if $\FAIL$ can be 
specified by priorities, and if that approach yields better-behaved 
semantics.

\subsection*{Acknowledgment}
The authors would like to thank Yoshinao Isobe for comments on an 
earlier draft of this paper and stimulating discussions.

\bibliographystyle{splncs03}
\bibliography{local}

\end{document}
